\RequirePackage{arydshln}
\documentclass[aps,twocolumn,nofootinbib,superscriptaddress,preprintnumbers,10pt,floatfix]{revtex4-1}

\usepackage[utf8]{inputenc}

\usepackage{float}
\usepackage{amsmath,amssymb}
\usepackage{dsfont} 
\usepackage{hyperref}
\usepackage{graphicx}
\usepackage{enumitem}
\usepackage{mathtools}
\usepackage{bbold}
\usepackage{multirow}
\usepackage{ytableau}
\usepackage{youngtab}
\usepackage{braket}
\usepackage{soul}
\usepackage{cancel}
\usepackage{xcolor}
\usepackage[normalem]{ulem}

\usepackage{lipsum}

\usepackage{colortbl} 
\definecolor{Gray}{gray}{0.95}
\definecolor{RGray}{gray}{0.90}
\definecolor{CGray}{gray}{0.92}
\definecolor{nicegreen}{rgb}{0.1,0.5,0.1}

\usepackage{arydshln}

\newcommand{\bs}{\mathbf}
\newcommand{\beq}{\begin{eqnarray}}
\newcommand{\eeq}{\end{eqnarray}}

\newcommand{\real}{{\sf I}\kern-.12em{\sf R}}
\newcommand{\comp}{{\sf I}\kern-.50em{\sf C}}
\newcommand{\unity}{{\sf I}\kern-.54em{\sf 1}}

\newcommand{\tr}{\mbox{Tr}}

\makeatletter
\g@addto@macro\bfseries{\boldmath}
\makeatother

\makeatletter
\renewcommand\paragraph{\@startsection{paragraph}{4}{\z@}%
                                    {3.25ex \@plus1ex \@minus.2ex}%
                                    {-1em}%
                                    {\normalfont\normalsize\bfseries}}
\makeatother

\allowdisplaybreaks
\begin{document}

\preprint{}
\preprint{}

\title{ Lattice QCD study of the $\chi_{c1}\to J/\psi\ \gamma$ decay }

\author{Damir Be\v{c}irevi\'c}
\email{damir.becirevic@ijclab.in2p3.fr}
\affiliation{IJCLab, P\^ole Th\'eorie (Bat.~210), CNRS/IN2P3 et Universit\'e Paris-Saclay, 91405 Orsay, France}
\author{Roberto Di~Palma} 
\email{roberto.dipalma@uniroma3.it}
\affiliation{Dipartimento di Matematica e Fisica, Universit\`a  Roma Tre and INFN, Sezione di Roma Tre, Via della Vasca Navale 84, I-00146 Rome, Italy}
\author{Roberto Frezzotti} 
\email{frezzotti@roma2.infn.it}
\affiliation{Dipartimento di Fisica and INFN, Universit\`a di Roma ``Tor Vergata",\\ Via della Ricerca Scientifica 1, I-00133 Roma, Italy}
\author{Giuseppe Gagliardi} 
\email{giuseppe.gagliardi@roma3.infn.it}
\affiliation{Dipartimento di Matematica e Fisica, Universit\`a  Roma Tre and INFN, Sezione di Roma Tre, Via della Vasca Navale 84, I-00146 Rome, Italy}

\author{Vittorio Lubicz} 
\email{vittorio.lubicz@roma3.infn.it}
\affiliation{Dipartimento di Matematica e Fisica, Universit\`a  Roma Tre and INFN, Sezione di Roma Tre, Via della Vasca Navale 84, I-00146 Rome, Italy}

\author{Francesco Sanfilippo}
\email{francesco.sanfilippo@infn.it}
\affiliation{Istituto Nazionale di Fisica Nucleare, Sezione di Roma Tre,\\ Via della Vasca Navale 84, I-00146 Rome, Italy}

\author{Nazario Tantalo} 
\email{nazario.tantalo@roma2.infn.it}
\affiliation{Dipartimento di Fisica and INFN, Universit\`a di Roma ``Tor Vergata",\\ Via della Ricerca Scientifica 1, I-00133 Roma, Italy}

\begin{abstract}
\vspace{5mm}
We present the results of our lattice QCD computation of the electric ($E_{1}$) and magnetic $(M_{2})$ form factors relevant to the $\chi_{c1}\to J/\psi\,\gamma$ decay by using the gauge field configurations produced by the Extended Twisted Mass Collaboration with $N_{f}=2+1+1$ dynamical Wilson-Clover twisted mass fermions at four different lattice spacings with physical dynamical $u$ , $d$, $s$ and $c$ quark masses (except for the coarsest lattice for which the lightest sea quark corresponds to a pion with $m_{\pi}\simeq 175~\mathrm{MeV}$). In the continuum limit, we obtain $\Gamma( \chi_{c1}\to J/\psi\ \gamma ) = 0.3265(79)~\mathrm{MeV}$, which agrees to $(1\div 2) \sigma$ with the experimental results and disagrees with a previous (unquenched) lattice QCD calculation. Our result for the magnetic quadrupole fractional transition amplitude, $a_{2} =M_{2}/\sqrt{E_{1}^{2}+M_{2}^{2}} = -0.0666(22)$, is in agreement with the experiment and  
represents an improvement by a factor of about $30$ with respect to the only existing (quenched) lattice QCD result. 
\vspace{3mm}
\end{abstract}

\maketitle

\allowdisplaybreaks

\section{Introduction}\label{sec:intro}
In recent years, the radiative decays of heavy quarkonia have attracted considerable phenomenological interest. The study of the so called electric dipole (E1) and magnetic dipole (M1) transitions, such as $h_{c(b)} \rightarrow \eta_{c(b)} \gamma$, $J/\psi \to \eta_{c}\gamma$ and $\Upsilon \to \eta_{b}\gamma$, can give further insight in the internal structure and dynamics of heavy quarkonium states, conveniently described by an appropriate effective field theory~\cite{Bodwin:1994jh,Brambilla:1999xf,Fleming:2000ib,Brambilla:2004jw}. 
Further motivation for studying these decay modes arises from the searches of a light (flavorless) pseudoscalar state, originating from beyond the Standard Model (BSM), that could either be a candidate for dark matter or act as a portal that mediates interaction between the Standard Model (SM) fermions and the dark matter constituents. In that respect the modes with $\eta_{c,b}$ in the final state are particularly interesting, as the observed state could be a mixture of the genuine $\eta_{c,b}$ (SM) and the light BSM pseudoscalar boson. To disentangle such a mixing, it is essential to keep the SM amplitude under theoretical control, which is where lattice QCD provides a key contribution. 
In that context we recently investigated the $1^{+-} \to 0^{-+} + \gamma$ decay, both in the case of charmonia and  bottomonia, which allowed us to predict for the first time the full width of the $h_{b}$ meson, $\Gamma(h_{b}) = (88 \pm 13)~\mathrm{keV}$, and to improve $\Gamma(h_{c}\to\eta_{c}\gamma)$ by more than a factor of two with respect to the previous lattice calculations.~\footnote{ We use the standard spectroscopic notation to label the states, $J^{PC}$, where $J$ stands for the quantum number associated to the angular momentum (spin), while $P$ and $C$ correspond to state's parity and  charge conjugation, respectively. }

In this paper, instead, we focus on the $1^{++}\to 1^{--}\gamma$ channel; the $\chi_{c1}\to J/\psi\ \gamma$ decay mode. That channel is interesting because it has been studied experimentally in great detail so that a comparison between theory (lattice QCD) and experiment, without involving weak interaction, can provide a fine verification of the validity of the precision lattice QCD computation of hadronic quantities. 
Regarding experiment, the decay width of this channel has been measured by the Fermilab collaborations~\cite{E760:1991qim,Andreotti:2005ts} and the current average is $\mathcal{B}(\chi_{c1} \to J/\psi\ \gamma) = (34.3\pm 1.3)\%$~\cite{PDG2024}. The charm factories at CLEO~\cite{CLEO:2009inp} and BES-III~\cite{BESIII:2017tsq} went a step beyond and made an accurate angular analysis of decays of the $\chi_{cJ}$ resonances, which then allowed them to extract the so called magnetic quadrupole fractional amplitude,
\begin{align}\label{eq:a2}
a_{2} = \frac{M_{2}}{\sqrt{E_{1}^{2} + M_{2}^{2}}}~,    
\end{align}
where $E_{1}\equiv E_1(0)$ and $M_{2}\equiv M_2(0)$ are respectively the electric and magnetic form factors describing $\chi_{c1}\to J/\psi \ \gamma$ (see Eq.~\eqref{eq:def_trans_ff} below). The results obtained by CLEO and BES-III, $a_{2}^{\rm CLEO} =-0.0626(67)$ and $a_{2}^{\rm BES-III} = -0.0740(47)$, have an accuracy of $O(10\%)$ and $O(6\%$), and differ from each other by $1.4\sigma$.

 Concerning the previous lattice QCD studies, this mode has been elaborated for the first time in Refs.~\cite{Dudek:2006ej,Dudek:2009kk} where the computation of the relevant form factors has been performed on a single (anisotropic) lattice in the quenched approximation. Note that they computed the form factors $E_{1}(q^2)$ and $M_{2}(q^2)$ at $q^2\neq 0$, and then extrapolated them to $E_{1}$ and $M_{2}$, corresponding to the on-shell photon ($q^2=0$). For that reason they could not provide a reliable determination of $a_{2}$. Instead, for the decay rate, which is dominated by the 
electric form factor $E_{1}(0)$, they reported $\Gamma(\chi_{c}\to J/\psi\ \gamma ) = 0.27(7)~{\rm MeV}$.
To date the only unquenched result for this decay has been reported in Ref.~\cite{Li:2022cfy}, obtained by using the gauge field configurations with $N_{f}=2$ twisted mass quarks at two lattice spacings. The decay rate was estimated to be $\Gamma(\chi_{c1}\to J/\psi\ \gamma) = (73\pm 20)~{\rm keV}$, thus much smaller than both the result of Ref.~\cite{Dudek:2009kk}, and the experiment $\Gamma(\chi_{c1}\to J/\psi\ \gamma)^\mathrm{exp} = 0.288(18)~\mathrm{MeV}$~\cite{PDG2024}.

In this work we seek to improve on the lattice results by performing a high precision calculation of both $E_{1}$ and $M_{2}$ form factors. In so doing we employ four different lattice spacings in the range $a\in [0.058-0.09]$~fm. A particularly subtle point, when dealing with twisted mass quarks, is to choose an appropriate  interpolating operator for the $\chi_{c1}$ state, i.e. the one that avoids mixing of the desired $J^{PC} = 1^{++}$ with unwanted $J^{PC}=1^{--}$ and/or $J^{PC} = 1^{+-}$ states, which are lighter and thus would obstruct the overlap with $\chi_{c1}$. Such a mixing is due to the parity breaking in the twisted mass QCD formulation and therefore it can show up even at zero three-momentum. We will show that this problem can be solved, and then through a high-statistics computation of the relevant two- and three-point correlation functions we determine both the electric and magnetic form factors with $O(1-3\%)$ accuracy. We anticipate our final results, namely
\begin{equation}
\begin{split}
\Gamma(\chi_{c1}\to  J/\psi\ \gamma) &= 0.3265(79)~{\rm MeV}, \\[2 ex]
a_{2}  &= -0.0666(22)\,.  
\end{split}
\end{equation}

The remainder of this paper is organized as follows. In Section~\ref{sec:num_details} we discuss the lattice setup employed for the determination of the two relevant form factors, focusing in particular on the choice of the interpolating operators adopted to isolate the $\chi_{c1}$ and the $J/\psi$. In Section~\ref{sec:num_results} we present the results of our analysis of the $J/\psi$ and $\chi_{c1}$ masses, as well as the results for the two form factors $E_{1}$ and $M_{2}$, including the discussion of all systematic effects including those associated with the correct isolation of the initial and final hadronic states and with the continuum-limit extrapolation. Finally, in Section~\ref{sec:comparison}, we compare our results with experimental measurements and existing theoretical calculations, and in Section~\ref{sec:conclusions} we present our conclusions and discuss directions for future improvements. \\

\section{$\chi_{c1}\to J/\psi$ Form Factors and Lattice Setup}
\label{sec:num_details}

We adopt the following decomposition of the transition matrix element describing the decay $\chi_{c1} \to J/\psi  \gamma^\ast $~\cite{Dudek:2006ej}:
\begin{widetext}
\begin{align}
\label{eq:def_trans_ff}
\langle J/\psi(k,\varepsilon) | J^{\mu}_{\rm em} | \chi_{c1}(p,\eta ) \rangle &= \frac{iQ_{c}}{4\sqrt{2}\  \Omega(q^2)} \epsilon^{\mu\nu\rho\sigma} (p
  - k)_\sigma \times  \Bigg\{ 2\left[ E_1(q^2) + M_{2}(q^{2})\right]  
  (\eta^\ast \cdot k) \, \varepsilon_{\nu} (p+k)_\rho + \nonumber \\
  &+ 2 \frac{m_{J/\psi}}{m_{\chi_{c1}}}\left[ E_{1}(q^{2})-M_{2}(q^{2})\right]  (\varepsilon \cdot p)
  \eta^\ast_{\nu}\, (p+k)_\rho + \frac{C_{1}(q^2)}{\sqrt{q^2}} \Big[ - 4
\Omega(q^2) \, \eta^\ast_{\nu} \varepsilon_{\rho}   \nonumber \\
&+  \big[ (m_{\chi_{c1}}^2-m_{J/\psi}^2 + q^2) ( \eta^\ast \cdot k)\; \varepsilon_{\nu} + (m_{\chi_{c1}}^2 - m_{J/\psi}^2 -q^2) (\varepsilon \cdot p) \; \eta^\ast_{\nu} \big]          \, (p+k)_\rho\Big]\Bigg\}\,,
\end{align}
\end{widetext}
where $J^{\mu}_{\rm em}$ is the electromagnetic current,
\begin{equation}
\label{eq:em_current}
    J^{\mu}_{\text{em}}(x) = \!\!\sum_{f=u,d,s,c} \!\! J_{f}^{\mu}(x) = \!\!\sum_{f=u,d,s,c} \!\! Q_f \, \bar{q}_f(x) \gamma^{\mu} q_f(x)\, ,
\end{equation}
with the sum running over all quark flavors, $ q_f(x) $ being the quark field, and $ Q_f $ its electric charge in units of $ e = \sqrt{4\pi \alpha_\mathrm{em}}$. In Eq.~\eqref{eq:def_trans_ff}, $ p $ and $ k $ are the momenta of $\chi_{c_1} $ and $J/ \psi$ respectively, while $\eta$ and $\varepsilon$ are their respective polarization vectors. Finally, $E_1(q^2)$, $M_2(q^2)$ and $C_1(q^2)$ are the form factors, functions of $q^2=(p-k)^2$, while the kinematic function 
 $\Omega(q^{2}) = (p\cdot k)^{2} - m_{J/\psi}^{2} m_{\chi_{c1}}^{2}$. The form factor $C_{1}\equiv C_{1}(0)$ does not contribute to the matrix element of the physical $\chi_{c1} \to J/\psi  \gamma$ and will not be discussed in the following. Note also  that we slightly modified the decomposition of Refs.~\cite{Dudek:2006ej,Dudek:2009kk,Li:2022cfy} in order to make both $E_{1}$ and $M_{2}$ dimensionless. The decay rate then reads:
\begin{align}
\label{eq:decay_rate_charm}
\Gamma\left( \chi_{c1} \to J/\psi\gamma \right) =  \frac{Q_{c}^{2}}{3}\alpha_{\rm em}\frac{ m_{\chi_{c1}}^{2} - m_{J/\psi}^{2}}{2
m_{\chi_{c1}}}\,\bigl( \, |E_{1}|^{2} + |M_{2}|^{2} \, \bigr)~.
\end{align}

In our computation we use the gauge field configurations produced by the Extended Twisted Mass Collaboration (ETMC) with $N_{f}=2+1+1$ dynamical Wilson-Clover twisted mass fermions which guarantees the automatic $O(a)$ improvement of parity-even observables~\cite{Frezzotti:2003ni,Frezzotti:2004wz}. Basic information regarding the four lattice ensembles used in this work is collected in Table~\ref{tab:simudetails}, and further details can be found in  Ref.~\cite{ExtendedTwistedMassCollaborationETMC:2024xdf}. For the reasons that will become clear later on, in this work we employ the mixed lattice action, a setup first introduced in Ref.~\cite{Frezzotti:2004wz} and then further described in Ref.~\cite{ExtendedTwistedMassCollaborationETMC:2024xdf}. In other words, in contrast to the sea quarks, the valence quark action is discretized with the so-called Osterwalder--Seiler (OS) regularization, namely, 
\begin{widetext}
\begin{align}
\label{eq:tm_action}
S = \sum_{f=u,d,s,c}\sum_{x} \bar{q}_{f}(x) \biggl[ \gamma_{\mu}\bar{\nabla}_{\mu}[U] -ir_{f}\gamma^{5}(W^{\rm cl}[U] + m_{\rm cr}) + m_{f}  \biggr] q_{f}(x)~,
\end{align}
\end{widetext}
where $W^{\rm cl}[U]$ is the Wilson-Clover term~\cite{Sheikholeslami:1985ij}, $m_{\rm cr}$ is the critical mass, $m_{f}$ the quark mass of the flavor $f$ (with $m_{u}=m_{d}=m_{l}$), and $r_{f}=\pm 1$ is the sign of the twisted-Wilson parameter for the flavor $f$. 
At each lattice spacing, the charm quark mass $m_{c}$ has been tuned to reproduce $m_{D_{s}}= 1968$~MeV. 
\begin{table}[t]
\begin{ruledtabular}
\begin{tabular}{lcccccc}
\textrm{ID} & $L/a$ & $a$ \textrm{fm} & $Z_{V}$ & $am_{c}$ & $N_{g}$ & $N_{s}$ \\
\colrule
\textrm{A48} & $48$ & 0.0907(5) &  0.68700(15) & $0.2620$ & $800$ & $140$ \\
\textrm{B64} & $64$ & 0.07948(11) &  0.706354(54) & $0.23157$  & $400$ & $32$ \\
\textrm{C80} & $80$ & 0.06819(14)  & 0.725440(33) & $0.19840$  & $640$ & $24$ \\
\textrm{D96} & $96$ & 0.056850(90) & 0.744132(31)   &  $0.16490$  &  $300$ & $35$ 
\end{tabular}
\end{ruledtabular}
\caption{\small\sl$N_{f}=2+1+1$ ETMC gauge ensembles used in this work. $a$ is the lattice spacing, $L/a$ spatial extent of the lattice, $am_{c}$ is the bare charm quark mass and $Z_V$ the vector current renormalization constant~\cite{ExtendedTwistedMassCollaborationETMC:2024xdf}. The number of gauge configurations $N_{g}$, and the number of (spin and color diluted) stochastic sources $N_{s}$ employed for the calculation of the relevant two- and three-point correlation functions are also given. With the exception of A48 (which corresponds to a pion mass $m_{\pi}\simeq 175~{\rm MeV}$), all the ensembles have been generated at physical values of the light, strange and charm quark masses. \label{tab:simudetails}}
\end{table}

We work in the $\chi_{c1}$ rest frame ($\mathbf{p}=0$). In order to ensure $q^2=0$ one needs to give $J/\psi$ a three-momentum 
\begin{align}
\label{eq:momentum_JPsi}
|\mathbf{k}| = \frac{ m_{\chi_{c1}}^{2} -m_{J/\psi}^{2} }{2m_{\chi_{c1}}} \simeq 389.4~\mathrm{MeV}\,. 
\end{align}
The form factors $E_{1}$ and $M_{2}$ can be extracted from the following three-point correlation functions:
\begin{widetext}
\begin{align}
\label{eq:three-point}
 C_{\text{3pt}}^{ijk}(t_{\chi}; t_{J}) = \sum_{\mathbf{x}, \mathbf{y},\mathbf{z}} e^{i\mathbf{k}(\mathbf{x}-\mathbf{y})} \langle 0 | \mathcal{O}_{J/\psi}^{i}(\mathbf{x}, 0) J^{j}_{\rm em}(\mathbf{y}, -t_{J}) \mathcal{O}_{\chi_{c1}}^{k \dagger}(\mathbf{z}, -t_{\chi}) | 0 \rangle~,
\end{align}
\end{widetext}
where $\mathcal{O}_{J/\psi}^{i}$ and $\mathcal{O}_{\chi_{c1}}^{k}$ are the interpolating operators of the $J/\psi$ and $\chi_{c1}$ mesons,
\begin{equation}
\label{eq:interpolators}
\begin{split}
\mathcal{O}_{J/\psi}^{i}(\mathbf{x},t) &= \sum_{\mathbf{y}}\bar{q}_{c}(\mathbf{x},t) G_{t}^{n}(\mathbf{x},\mathbf{y}) \gamma^{i} q_{c}(\mathbf{y},t) \,,  \\ 
\mathcal{O}_{\chi_{c1}}^{k}(\mathbf{x},t) &= \sum_{\mathbf{y}}\bar{q}_{c}(\mathbf{x},t) G_{t}^{n}(\mathbf{x},\mathbf{y}) \gamma^{k}\gamma^{5} q_{c}(\mathbf{y},t)  \,,
\end{split}
\end{equation}
where 
\begin{align}
\label{eq:G_def}
G_{t}(\mathbf{x},\mathbf{y}) = \frac{1}{1+ 6\kappa}\bigl[ \delta_{\mathbf{x},\mathbf{y}} + \kappa H_{t}(\mathbf{x},\mathbf{y})    \bigr]\,,
\end{align}
and $H_{t}(\mathbf{x},\mathbf{y})$ is the Gaussian smearing operator, 
\begin{align}
\label{eq:H}
H_{t}(\mathbf{x}, \mathbf{y}) = \sum_{\mu=1}^{3}\bigl[ U^{\star}_{\mu}(\mathbf{x},t)\delta_{\mathbf{x}+\hat{\mu},\mathbf{y}} + U^{\star\dagger}_{\mu}(\mathbf{x}-\hat{\mu},t)\delta_{\mathbf{x}-\hat{\mu},\mathbf{y}}    \bigr]\,,
\end{align}
with $U^{\star}_{\mu}(x)$ being the so-called APE-smeared links, cf.~\cite{Becirevic:2012dc}. As in our recent work~\cite{Becirevic:2025ocx},
we used the smearing parameter $\kappa=0.4$, and fix on each ensemble the number of steps $n$ in Eq.~\eqref{eq:interpolators} to have a smearing radius $r_{0}= a\sqrt{n}/\sqrt{\kappa^{-1}+6} \simeq 0.15~{\rm fm}$, for both $J/\psi$ and $\chi_{c1}$~\cite{DiPalma:2024jsp}. The spatial momentum $\mathbf{k}$ is injected along the third spatial direction, $\mathbf{k} = (0,0,|\mathbf{k}|)$. With this choice, the interpolating operators in Eq~\eqref{eq:interpolators} are such that $O_{\chi_{c1}}^{k,\dagger}$ creates a $\chi_{c1}$ with the polarization vector $\eta^{\mu}_{k} = \delta^{\mu}_{k}$, for three polarization states $k=1,2,3$, while $O_{J/\psi}^{i}$ annihilates a $J/\psi$ with $\varepsilon^{\mu}_{i} = \delta^{\mu}_{i}$ for $i=1,2$, and with $\varepsilon_{3}  = ( \alpha, 0, 0,\sqrt{1+\alpha^{2}})$, for the longitudinal polarization $i=3$, where $\alpha= |\bs{k}|/m_{J/\psi}$.

The Wick contractions in Eq.~\eqref{eq:three-point} give rise to the quark-connected and disconnected contributions. Just as in our recent work~\cite{Becirevic:2025ocx}, here we also focus on the evaluation of the dominant connected diagram, leaving the evaluation of the disconnected contributions for future work.  Clearly, only the charm quark component of the electromagnetic current ($J_{c}^{\mu}$) contributes to the connected part of $C_{\rm 3pt}^{ijk}(t_{\chi};t_{J})$. 
We use twisted boundary conditions to tune the spatial momentum $\mathbf{k}$ to the value given in Eq.~\eqref{eq:momentum_JPsi}. This is implemented by twisting the gauge links $U_{\mu}(x)$ on which one of the charm quark propagators is computed, namely~\cite{deDivitiis:2004kq,Bedaque:2004kc,Sachrajda:2004mi} 
\begin{align}
\label{eq:twisted_gauge}
U_\mu(x) \rightarrow U^\theta_\mu (x) = e^{i a\theta_{\mu} / L} U_\mu(x), \quad  \theta_\mu = (0, \vec{\theta})\,,
\end{align}
with the twisting-angle $\vec{\theta}$ set to
\begin{align}
\label{eq:theta}
\vec{\theta}=( 0, 0 , \theta_{z}^{c}) , \qquad \theta_{z}^{c} = \frac{L}{\pi}|\bs{k}|~. 
\end{align}
We denote by $S_{c}(x,y)$ and $S_{c}^{\theta}(x,y)$ the charm quark propagators evaluated on the background field configurations corresponding to $U_{\mu}(x)$ and $U_{\mu}^{\theta}(x)$, respectively, so that $C_{\rm 3pt}^{ijk}(t_{\chi}; t_{J})$ can be written as:~\footnote{To simplify the expressions we consider the case of local interpolating operators, i.e. $\kappa=0$. In the computations, however, we implemented the smearing in a way discussed in the text.} 
\begin{align}
\label{eq:3pt_explicit}
\hspace*{-3mm}C_{\rm 3pt}^{ijk}(t_{\chi}, t_{J}) = 2 \sum_{\mathbf{x},\mathbf{y},\mathbf{z}} \langle \tr \left[ S_{c}(x, z) \gamma^{k}\gamma^{5} S_{c}(z,y) \gamma^{j} S_{c}^{\theta}(y, x) \gamma^{i} \right]     \rangle\,,
\end{align}
where $x=(\mathbf{x},0)$, $y=(\mathbf{y}, -t_{J})$, $z=(\mathbf{z}, -t_{\chi})$. The trace $\tr\left[ \dots \right]$ is taken over the color and Dirac indices, and $\langle \dots \rangle$ indicates the average over the $\rm{SU}(3)$ gauge field configurations $U_{\mu}(x)$. The factor of two in Eq.~\eqref{eq:3pt_explicit} accounts for the contribution of the charge conjugated diagram. The sum over $\mathbf{x}$ is stochastically evaluated by inverting the Dirac charm quark operator on a number $N_{s}$ of (spin and color diluted) spatial stochastic sources placed at time $t=0$. The backward propagator $S_{c}(x,z)$ is then obtained from $S_{c}(z,x)$ by using $\gamma^{5}$-hermiticity.   
\subsection{Avoiding the spurious mixing }
We now discuss how the charm quark bilinears $\mathcal{O}_{J/\psi}$, $\mathcal{O}_{\chi_{c1}}$, and the current $J_{c}^\mu$, are discretized on the lattice. When evaluating the connected diagrams using the twisted mass (TM) action, there are different options when it comes to discretizing the quark bilinears and they differ in the choice of the  Wilson parameter of the quark field, $r_{+}=\pm 1$, and of the antiquark field, $r_{-}= \pm 1$. They can be chosen to be opposite ($r_{+}=-r_{-}$) or equal ($r_{+} = r_{-}$). The two regularizations are known as the TM and OS regularizations, respectively. The results obtained using different combinations of TM and OS bilinears differ only by $O(a^{2})$ effects. In Ref.~\cite{Becirevic:2025ocx}, we have shown that the TM regularization of the $h_{c}$ interpolating operator ($J^{PC} = 1^{+-}$) is protected from mixing with $J^{PC} = 1^{--}$ and $1^{++}$ states. This is so because in the TM regularization the product between parity $P$ and G-parity $G$, is an exact lattice symmetry, which is then sufficient to exclude the aforementioned mixing. In the case at hand, instead,  the exact $P\times G$ symmetry is not sufficient to exclude the mixing between $1^{++}$ and $1^{--}$ states. Indeed, when using the TM regularization the corresponding interpolating operator $O_{\chi_{c1}}^{k\dagger}$ also creates vector states, as we numerically illustrate in Sect.~\ref{sec:num_results}. In the OS regularization ($r_{+} = r_{-})$, however, one can use the exact invariance under charge conjugation $C$, which in our case forbids the creation of states $J^{PC} = 1^{--}$ and $J^{PC} = 1^{+-}$ when using the interpolating operators $O_{\chi_{c1}}^{k\dagger}$, thus representing a favorable regularization.

On the basis of the considerations made above, we therefore employ the OS regularization both for interpolating fields $O_{\chi_{c1}}$, $O_{J/\psi}$, and for the electromagnetic current.~\footnote{When using the OS regularization for the $J/\psi$ interpolating operator, it is possible that $J^{PC}=1^{+-}$ states are created, however, being heavier than the $J/\psi$ they do not represent an obstacle to our computation.}  With this choice, the renormalization constant of the local electromagnetic current is given by $Z_{V}(g_{0}^{2})$, the values of which are given in Tab.~\ref{tab:simudetails}. For a comprehensive discussion on bilinear interpolators in twisted mass lattice QCD see Ref.~\cite{Petry:2008rt}.

\subsection{Extraction of the form factors $E_{1}$ and $M_{2}$ }
\label{sec:extr_form_factor_charm}
We first need to consider the two point functions of interpolating fields $\mathcal{O}_{J/\psi}^{i}$ and $\mathcal{O}^{j}_{\chi_{c1}}$, respectively, in order to extract their mass/energy and their couplings:
\begin{equation}
\label{eq:2pt_def}
\begin{split}
    C_{2,J/\psi}^{i}(t) &= \sum_{\mathbf{x}} e^{i\mathbf{k}\mathbf{x}} \langle 0 | \mathcal{O}_{J/\psi}^{i}(\mathbf{x}, t) \mathcal{O}_{J/\psi}^{i \dagger}(\mathbf{0}, 0) | 0 \rangle\,, \\
    C_{2,\chi_{c1}}^{k}(t) &= \sum_{\mathbf{x}} \langle 0 | \mathcal{O}_{\chi_{c1}}^{k}(\mathbf{x}, t) \mathcal{O}_{\chi_{c1}}^{k \dagger}(\mathbf{0}, 0) | 0 \rangle\,,
\end{split}
\end{equation}
in which, as before, we neglect the Zweig suppressed disconnected contributions. 
In the limit of large Euclidean time $ t $, the two point correlation functions behave as:
\begin{equation}
\label{eq:2pt_asympt}
\begin{split}
    C_{2,J/\psi}^{i}(t) &= \frac{|Z_{J/\psi}^{i}|^2}{2 E_{J/\psi}} \left( e^{-E_{J/\psi} t} + e^{-E_{J/Psi} (T-t)}   \right) + \ldots ,  \\
     C_{2,\chi_{c1}}^{k}(t) &= \frac{|Z_{\chi_{c1}}^{k}|^2}{2 m_{\chi_{c1}}} \left( e^{-m_{\chi_{c1}} t} + e^{-m_{\chi_{c1}} (T-t)}  \right ) + \ldots~,
\end{split}
\end{equation}
where $ m_{\chi_{c1}} $ and $ E_{J/\psi} $ are the mass and energy of the corresponding ground states, and 
\begin{equation}
\begin{split}
Z_{J/\psi}^{i} &= \langle 0 | \mathcal{O}_{J/\psi}^{i} | J/\psi(k,\varepsilon_{i}) \rangle\,, \\
Z_{\chi_{c1}}^{k} &= \langle \chi_{c1}(p,\eta_{k} ) | \mathcal{O}^{k\dagger}_{\chi_{c1}} | 0 \rangle\,.
\end{split}
\end{equation}
The ellipses in Eq.~\eqref{eq:2pt_asympt} denote contributions that vanish in the large time limit, $0 \ll t \ll T$, and are further suppressed by the smearing, cf. Eq.~\eqref{eq:G_def}. We also calculated $C_{2,J/\psi}^{i}(t)$ with $\mathbf{k}=\bf{0}$, in order to extract $m_{J/\psi}$. Clearly, with our choice of kinematics we have: $Z_{\chi_{c1}}^{1}=Z_{\chi_{c1}}^{2}=Z_{\chi_{c1}}^{3}$, and $Z_{J/\psi}^{1} = Z_{J/\psi}^{2}$. 

It is then straightforward to use Eq.~\eqref{eq:def_trans_ff} and combine the above quantities into 
\begin{equation}
\label{eq:FF_estimators}
\begin{split}
\overline{E}_{1}(t_{\chi}; t_{J}) =& \frac{i Z_{V}(g_0^2)}{\sqrt{2}Q_{c}}  4 \, E_{J/\psi}  \  e^{E_{J/\psi}t_{J}} \  e^{m_{\chi_{c1}}(t_{\chi}-t_{J})} \\
&\quad \times \left[ \frac{C_{\rm 3pt}^{213}(t_{\chi};t_{J})}{{Z^{2}_{J/\psi} Z^{3}_{\chi_{c1}}}} - \frac{C_{\rm 3pt}^{312}(t_{\chi};t_{J})}{Z^{3}_{J/\psi} Z^{2}_{\chi_{c1}}} \right]\,,  \\[10pt]
\overline{M}_{2}(t_{\chi}; t_{J}) =&\frac{i Z_{V}(g_0^2)}{\sqrt{2}Q_{c}}  4 \,  E_{J/\psi}  \  e^{E_{J/\psi}t_{J}} \  e^{m_{\chi_{c1}}(t_{\chi}-t_{J})} \\
& \quad \times \left[ \frac{C_{\rm 3pt}^{213}(t_{\chi};t_{J})}{{Z^{2}_{J/\psi} Z^{3}_{\chi_{c1}}}} + \frac{C_{\rm 3pt}^{312}(t_{\chi};t_{J})}{Z^{3}_{J/\psi} Z^{2}_{\chi_{c1}}} \right]\,,
\end{split}
\end{equation}
which, for large Euclidean time separations, yield the desired form factors $E_{1}$ and $M_{2}$, namely, 
\begin{equation}
\begin{split}
\overline{E}_{1}(t_{\chi}; t_{J}) \xrightarrow{\begin{array}{c}
\scriptstyle t_{J} \to \infty \\
\scriptstyle t_{\chi} - t_{J} \to \infty
\end{array}}
            E_{1}\,~, \\
\overline{M}_{2}(t_{\chi}; t_{J}) \xrightarrow{\begin{array}{c}
\scriptstyle t_{J} \to \infty \\
\scriptstyle t_{\chi} - t_{J} \to \infty
\end{array}}
            M_{2}\,. 
\end{split}
\end{equation}
Note that in $C_{\rm 3pt}^{ijk}(t_{\chi};t_{J})$ the interpolating field of $J/\psi$ is placed at $t=0$, the charm quark component of the electromagnetic current is then inserted into a charm quark line at a fixed $-t_{J} < 0$, while the time $-t_{\chi} \ll -t_{J}$, at which the $\chi_{c1}$ is created, corresponds to the \textit{sink} of the correlation function is free. This allows us to carefully monitor the onset of dominance of the $\chi_{c 1}$ state. Since in our computational setup the time $t_{J}$ is fixed, we must choose it to be large enough to be able to isolate the $J/\psi$ state. The advantage of placing the interpolator corresponding to $J/\psi$ at the source of the correlation function is that $t_{J}$ can be chosen reasonably large since the signal-to-noise ratio (S/N) of $C_{\rm 3pt}^{ijk}(t_{\chi}; t_{J})$ for fixed $|t_{\chi}-t_{J}|$ decreases only as $e^{-(E_{J/\psi}-m_{\eta_{c}})} \simeq e^{-\Delta_{c}t}$, where $\Delta_{c}\simeq 113~{\rm MeV}$ is the hyperfine splitting. As will be detailed in the next subsection, we have considered two different values of $t_{J} \simeq 1.6, 2.4~\mathrm{fm}$, and observed only tiny differences in our results that we nevertheless include in the final systematic error.

\section{Numerical results }
\label{sec:num_results}
\begin{figure}[]
\centering
\includegraphics[scale=0.36]{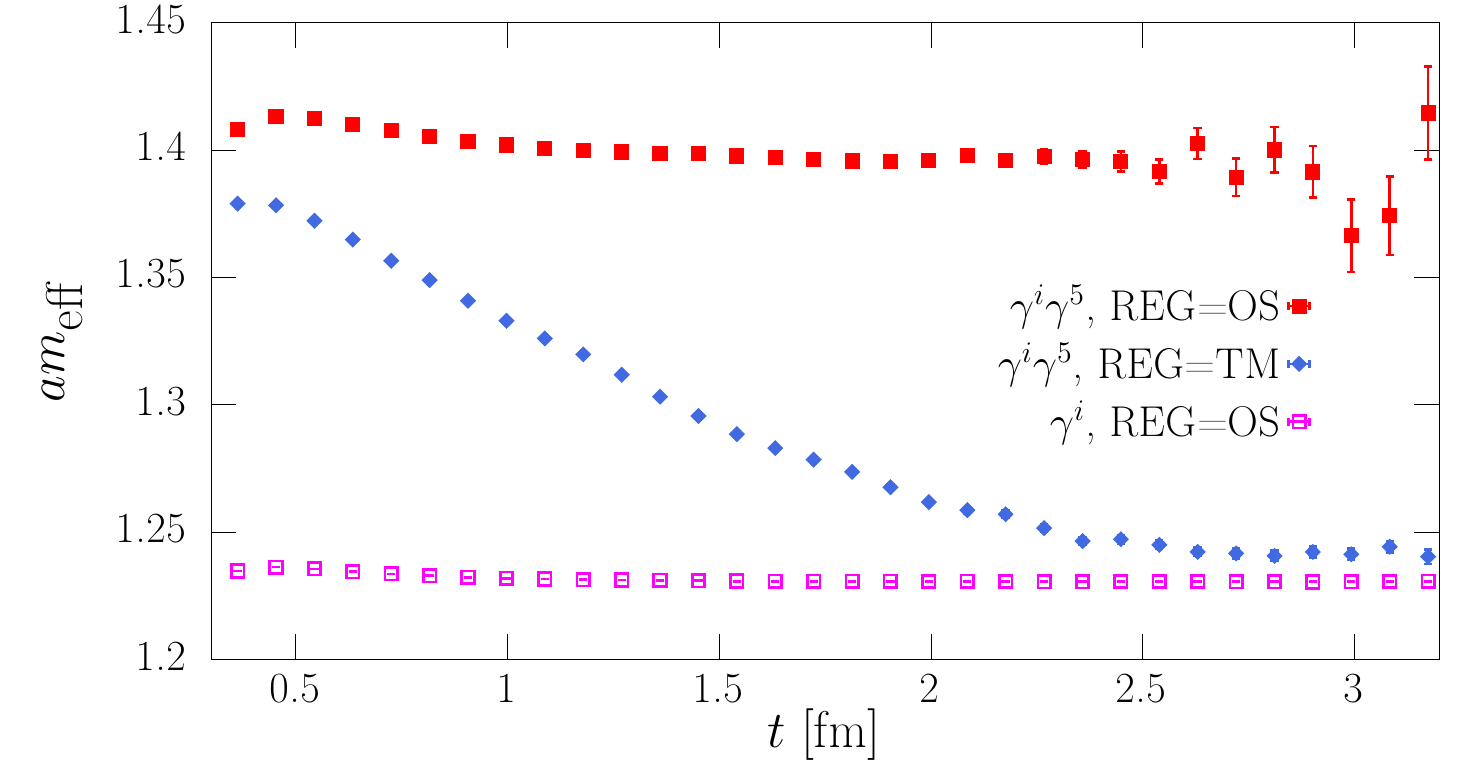}
\caption{\small\sl Effective mass extracted from $C_{2,\chi_{c1}}^{k}(t)$ in the TM (blue diamonds) and OS (red squares) regularizations, along with the one extracted from  $C_{2,J/\psi}^{i}(t)$ (with $|\bs{k}|=0$) in the OS regularization (magenta squares). The data refer to the ensemble B64. \label{fig:comp_TM_OS}}
\end{figure}
In Fig.~\ref{fig:comp_TM_OS} we show the effective mass of the two-point function $C_{2,\chi_{c1}}^{k}(t)$ regularized in both the TM (blue diamond) and OS (red squares) regularizations, as well as the effective mass of $C_{2,J/\psi}^{i}(t)$ (for $|\bs{k}|=0$) in the OS regularization (magenta squares). The results correspond to our ensemble B64. We clearly see that with the TM regularization the function $C_{2,\chi_{c1}}^{k}(t)$ exhibits mixing with $J/\psi$. However, due to the exact $C$-symmetry, such spurious mixing is absent when using the OS regularization which is why, in this work, we use that regularization for the valence quarks. 

\begin{figure}[]
\centering
\includegraphics[scale=0.35]{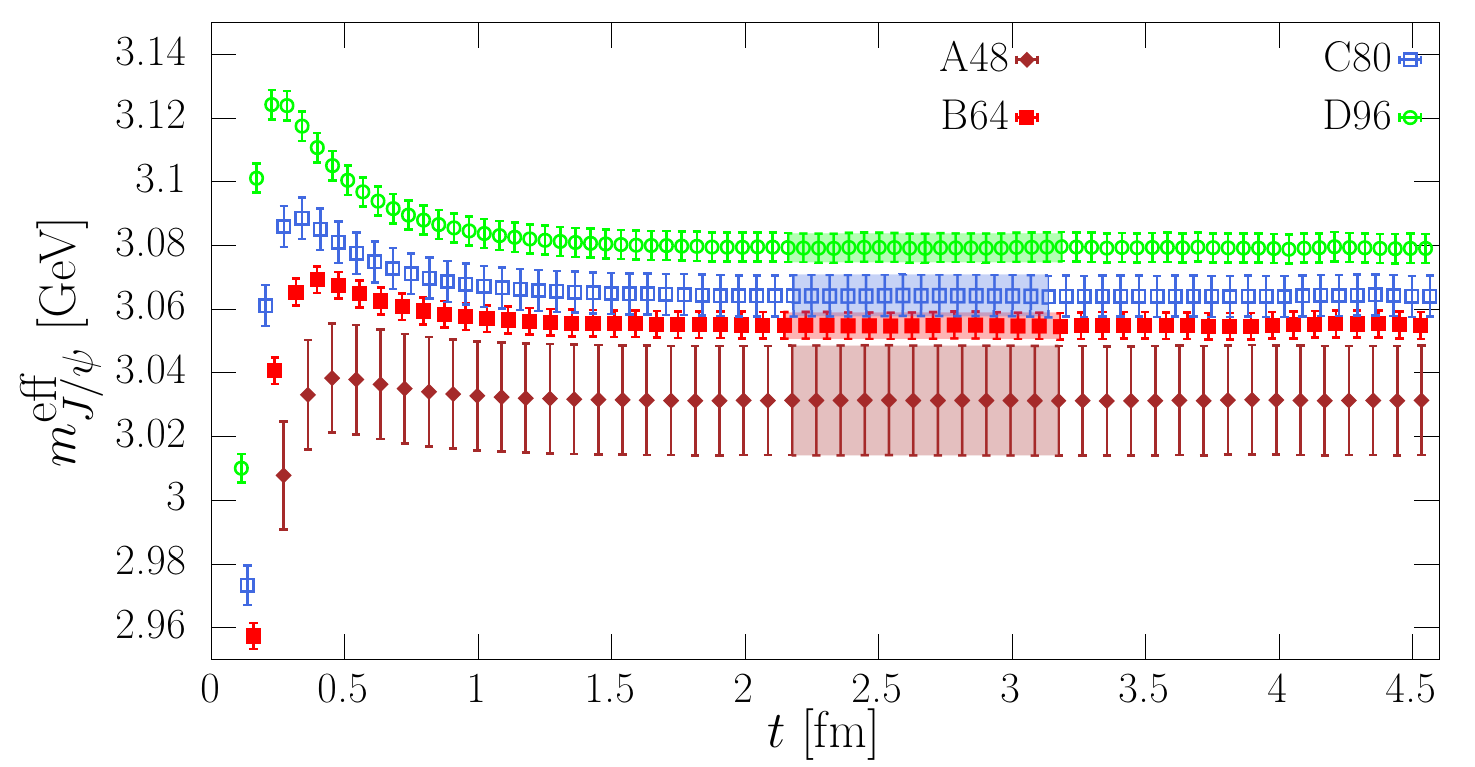}
\includegraphics[scale=0.35]{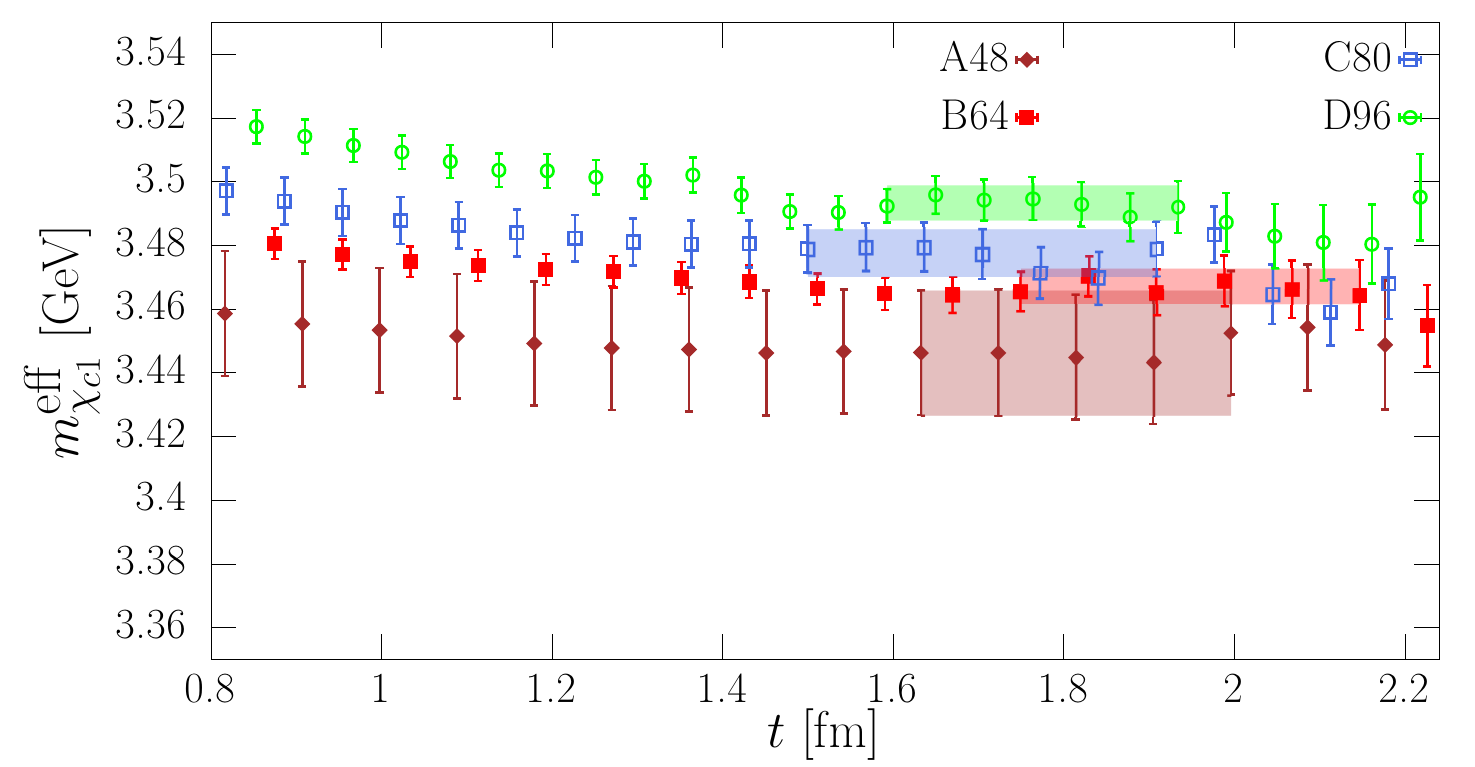}
\caption{\small\sl Effective mass of $J/\psi$ and $\chi_{c1}$ for all four gauge ensembles used in this work. The colored bands correspond to our estimate of mass on each ensemble. \label{fig:chi_Jpsi_masses}}
\end{figure}

\begin{figure}
\centering
\includegraphics[scale=0.35]{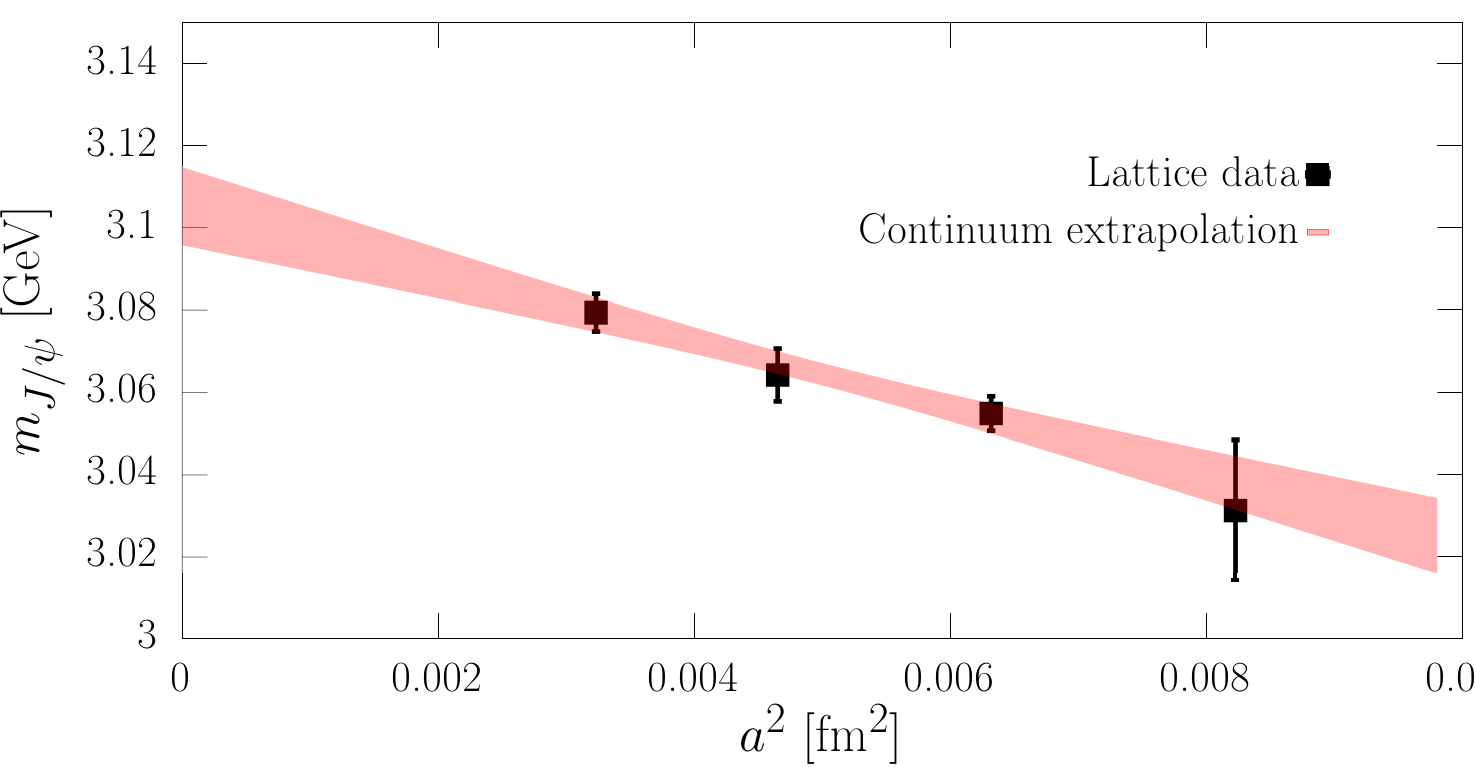}
\includegraphics[scale=0.35]{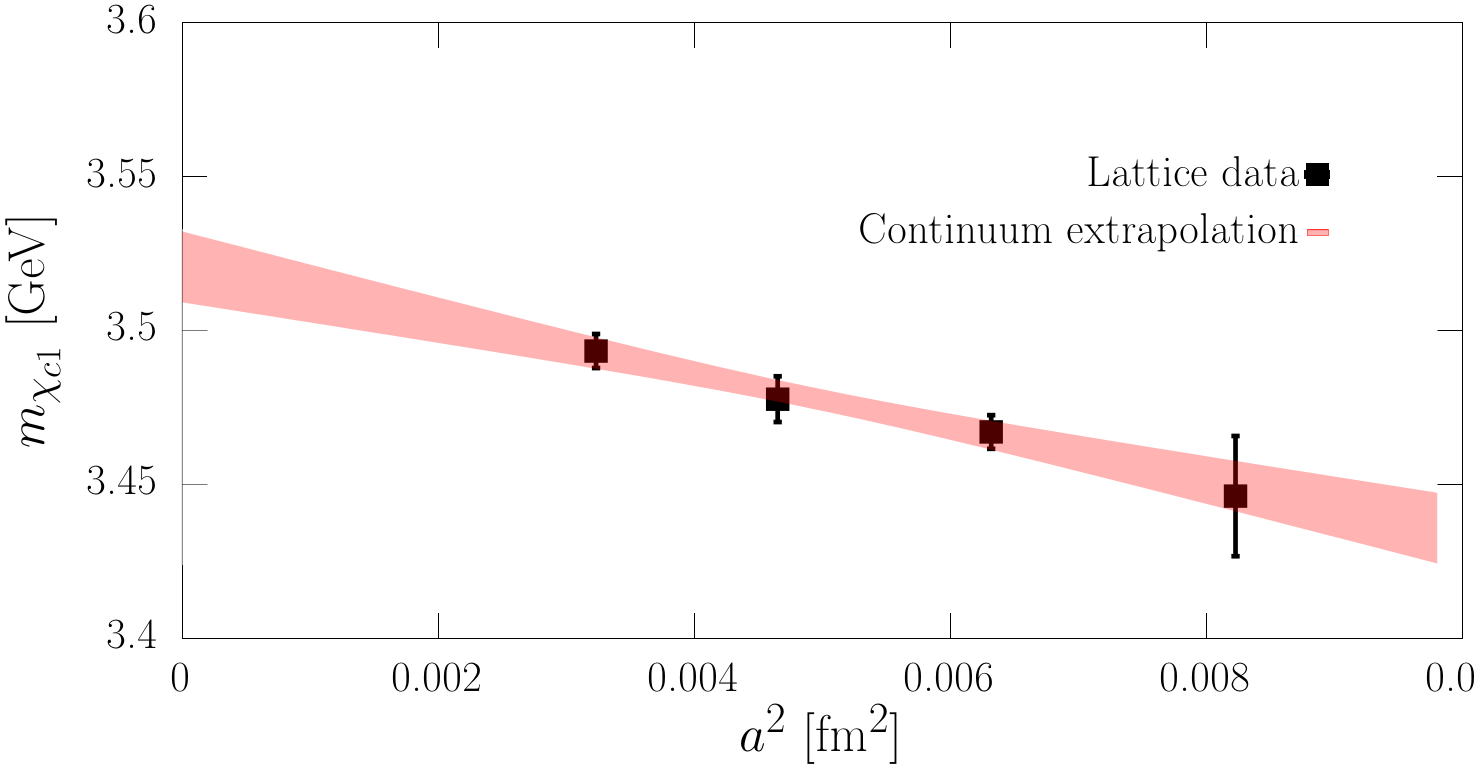}
\caption{\small\sl Extrapolation of $m_{J/\psi}$ and $m_{\chi_{c 1}}$ to the continuum limit ($a\to 0$). The colored bands correspond to the best fit obtained after performing a linear extrapolation in $a^{2}$. The reduced $\chi^{2}$ of the two fits is below one. \label{fig:chi_JPsi_extr}}
\end{figure}

In Fig.~\ref{fig:chi_Jpsi_masses} we show the effective mass plots for $J/\psi$ and $\chi_{c1}$ mesons, for all four of our lattices. 
We should emphasize that the charm quark mass has been fixed by $m_{D_{s}} =1968~\mathrm{MeV}$ at each value of the lattice spacing, and due to the finite lattice artefacts the values of $m_{J/\psi}$ and $m_{\chi_{c1}}$ are not the same for all ensembles. Owing to the fact that the S/N ratio of $C_{2,J/\psi}^{i}(t)$ decreases only as $e^{-\Delta_{c}t}$, the quality of the plateaus for $J/\psi$ are excellent. In the case of $C_{2,\chi_{c1}}(t)$ the S/N ratio decreases faster but since we use the smeared sources, as well as a large number of gauge configurations and stochastic sources, we are still able to obtain the long plateaus and extract $m_{\chi_{c1}}$ with a very good accuracy. 
Note that a larger error for the A48 ensemble in Fig.~\ref{fig:chi_Jpsi_masses} is due to uncertainty on the lattice spacing of that ensemble ($\simeq 0.6\%$). The extrapolation to the continuum of the lattice results for $m_{J/\psi}$ and $m_{\chi_{c1}}$ is shown in Fig.~\ref{fig:chi_JPsi_extr}. 
Since there are no signals of $a^{4}$ effects in the data, the continuum limit is reached through a simple linear fit in $a^{2}$ and for our final results we have:
\begin{equation}
\label{eq:etac_hc_cont_val}
\begin{split}
m_{J/\psi} &= 3.1052(95)~\mathrm{GeV}\,,\\[1.5 ex]
 \qquad m_{\chi_{c1}} &= 3.521(12)~\mathrm{GeV}\,,
\end{split}
\end{equation}
in good agreement with the current experimental values, $m_{J/\psi}^\mathrm{exp} = 3.096900(6)~\mathrm{GeV}$ and $m_{\chi_{c1}}^\mathrm{exp} = 3.51067(5)~\mathrm{GeV}$~\cite{PDG2024}. These results also suggest that the impact of neglected disconnected (annihilation) contributions is indeed small, 
at most of the order of our statistical uncertainty which is about $(0.3\div 0.4)\%$ for both masses. 

We now focus on the determination of the form factors $E_{1}$ and $M_{2}$. At finite lattice spacing we tune the three-momentum $\bs{k}$ to the value indicated in Eq.~\eqref{eq:momentum_JPsi} but by using the experimental values of $m_{J/\psi} $ and $ m_{\chi_{c1}}$, and not those obtained on the lattice at fixed lattice spacing. 
The two choices merely amount to a redefinition of the $O(a^{2})$ effects because, as we have just shown, our values of the masses in the continuum limit are in full agreement with the experimental ones. In practice, the use of the ensemble values of the masses in Eq.~(\ref{eq:momentum_JPsi}) would have produced on the coarsest lattice spacing a variation of the three-momentum $\bs{k}$ by less than $1\%$, which besides being only an $O(a^{2})$ effect, is also completely negligible at the current level of accuracy.

\begin{figure}
\centering
\includegraphics[scale=0.35]{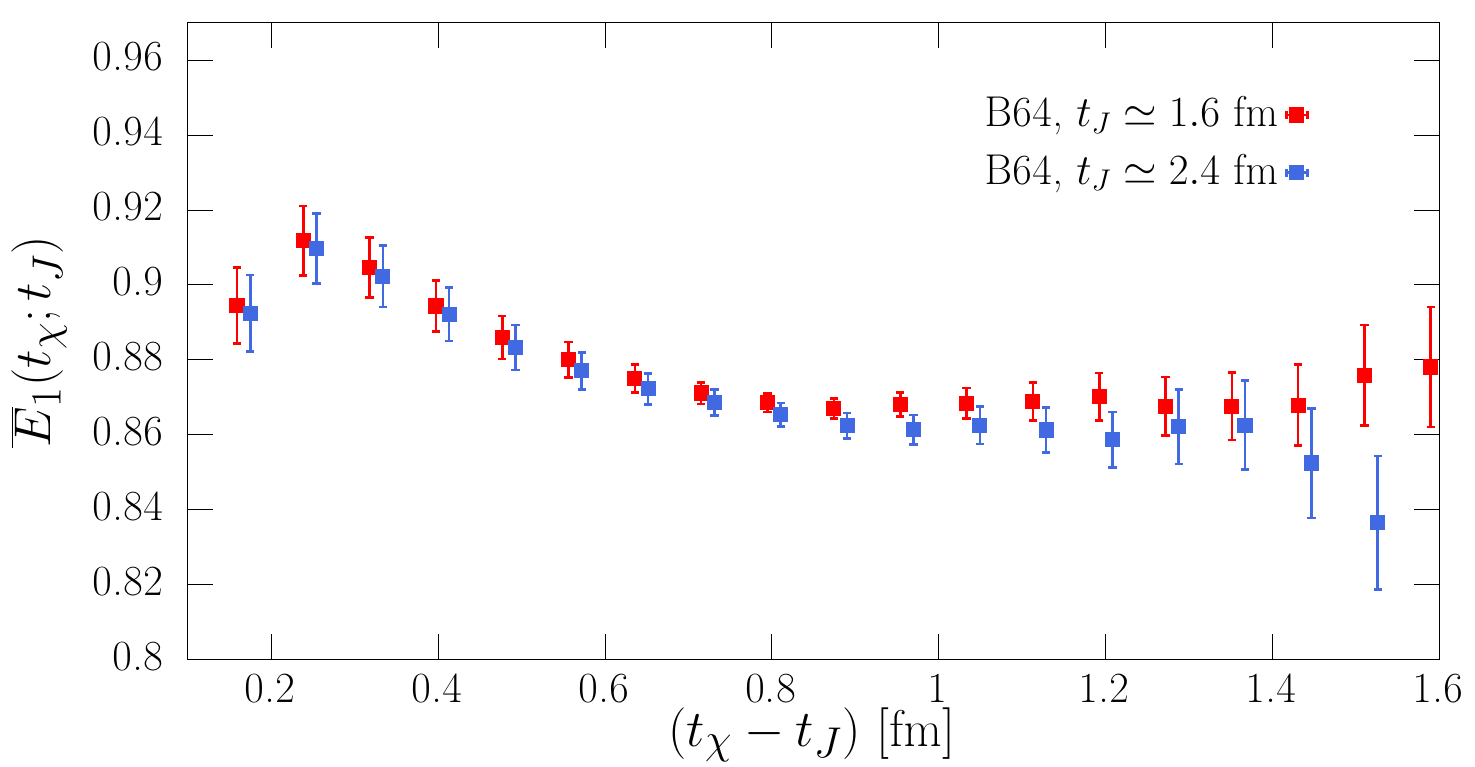}
\includegraphics[scale=0.35]{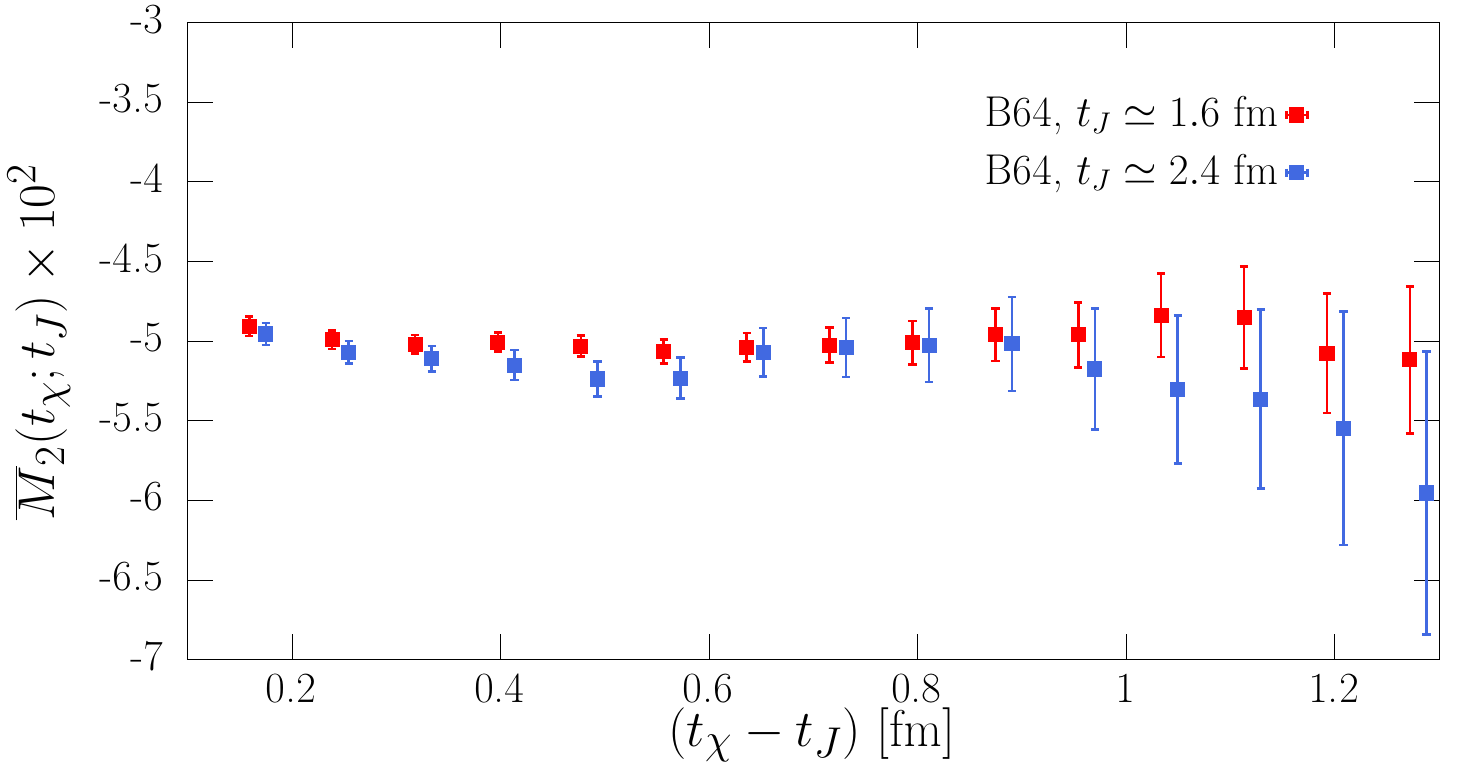}
\caption{\small\sl Comparison between $\overline{E}_{1}(t_{\chi}, t_{J})$ (left plot) and $\overline{M}_{2}(t_{\chi}, t_{J})$ (right plot) with $t_{J} \simeq 1.6~\mathrm{fm}$ (red) and with $t_{J} \simeq 2.4~\mathrm{fm}$ (blue) on the B64 ensemble. The datapoints at $t_{J}\simeq 2.4~\mathrm{fm}$ have been slightly shifted horizontally for easier comparison. \label{fig:3pt_test_B64}}
\end{figure}

As already anticipated in the previous Section, we keep the time $-t_{J}$ of the transition operator in $C_{\rm 3pt}^{ijk}(t_{\chi}; t_{J})$ fixed, while $t_{\chi}$ runs over the lattice. It is thus important to choose sufficiently large $t_{J}$ in order to ensure the dominance of $J/\psi$.  We tested that issue on the B64 ensemble by computing $C_{\rm 3pt}^{ijk}(t_{\chi}; t_{J})$ for two different $t_{J} \sim 1.6, 2.4~\mathrm{fm}$. In Fig.~\ref{fig:3pt_test_B64} we compare the corresponding $\overline{E}_{1}(t_{\chi}; t_{J})$ and $\overline{M}_{2}(t_{\chi}; t_{J})$, cf. Eq.~\eqref{eq:FF_estimators}.
The two results are in good agreement, and only in the case of $E_{1}$ we do see an $O(1\sigma)$ difference between the results obtained at the two values of $t_{J}$. On the basis of these observations we decided to fix $t_{J} \simeq 1.7~\mathrm{fm}$ for all other ensembles, and include, before the continuum extrapolation, a systematic uncertainty $\Delta E_{1}$ and $\Delta M_{2}$ to all our results, estimated as:
\begin{equation}
\begin{split}
\Sigma_{F} =  | F(t_{J}\simeq 2.4~{\rm fm}) - F(t_{J} \simeq 1.6~{\rm fm}) |~, \\
\Delta F = \Sigma_{F} \cdot  {\rm erf}\left(\frac{\Sigma_{F}}{\sqrt{2} \sigma_{\Sigma_{F}}}\right)~,\quad F=E_{1},M_{2}~,
\end{split}
\end{equation}
where $\sigma_{\Sigma_{F}}$ is the statistical error of $\Sigma_{F}$ and  ${\rm erf}(x)$ is the error function. The quantities $\Delta E_{1}$ and $\Delta M_{2}$ are thus the observed spreads, weighted by the probability that they are not due to statistical fluctuations.

\begin{figure}[t]
\centering
\includegraphics[scale=0.35]{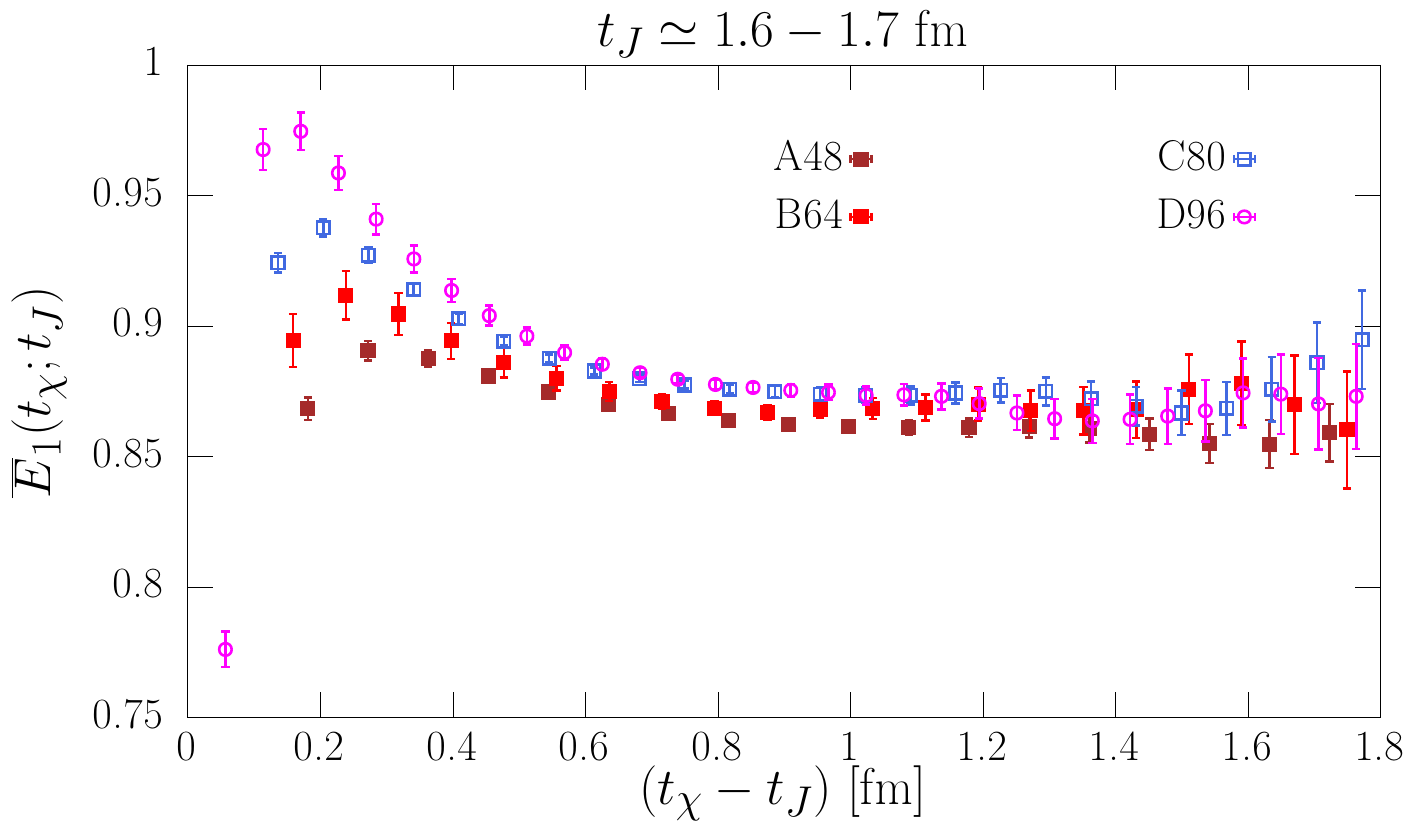} \\
\includegraphics[scale=0.35]{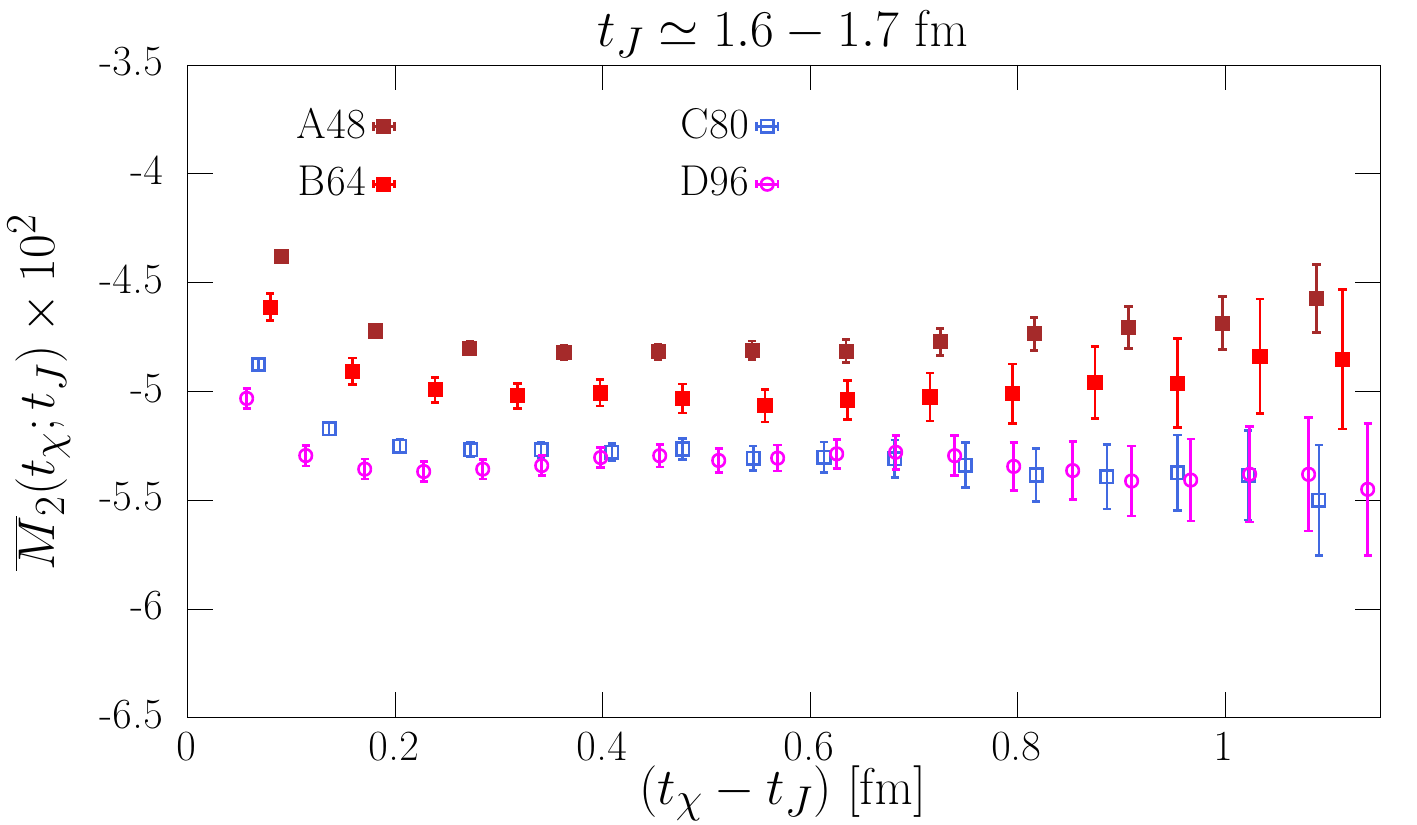}
\caption{\small\sl The functions $\overline{E}_{1}(t_{\chi}; t_{J})$ (left) and $\overline{M}_{2}(t_{\chi}; t_{J})$ (right)  defined in Eq.~\eqref{eq:FF_estimators}, as obtained for all lattice ensembles, cf. Tab.~\ref{tab:simudetails}, and for $t_{J}$ fixed as discussed in the text. \label{fig:F_all_beta}}
\end{figure}

In Fig.~\ref{fig:F_all_beta} we show $\overline{E}_{1}(t_{\chi}; t_{J})$ and $\overline{M}_{2}(t_{\chi}; t_{J})$ for all our lattices.~\footnote{For the A48 and D96 ensemble we have averaged the estimators for the form factors in Eq.~(\ref{eq:FF_estimators}) with the equivalent estimators obtained using the second component $J_{\rm em}^{2}$ of the electromagnetic current.} For the electric form factor $E_{1}$ the cut-off effects are small, of the order of a few percent, in addition to the statistical uncertainties that are $O(1\div 2\%)$). For the magnetic form factor $M_{2}$, instead, we observe the larger cut-off effects $O(10-15\%)$ that dominate the tiny statistical errors.  
Note also that the onset of the plateau region occurs already at small times, $\sim 0.5~{\rm fm}$. The form factors are then merely obtained from the fit to a constant.  

\begin{table}[t]
\begin{ruledtabular}
\begin{tabular}{lcccc}
$a~[{\rm fm}]$ & $0.0907\,(5)$ & $0.07948\, (11)$  & $0.06819\,(14)$    &  $0.056850\,(90)$    \\
\colrule
$E_{1}$  & $0.8609(60)$ & $0.8686(72)$ & $0.8738(64)$ &  $0.8699(74)$ \\
$-M_{2}$  & $0.04741(72)$ & $0.0502(11)$ & $0.0535(11)$ &  $0.0535(12)$
\end{tabular}
\end{ruledtabular}
\caption{\small\sl Values of the form factors $E_{1}$ and $M_{2}$ for each gauge ensemble used in this study. \label{tab:FF}}
\end{table}

Our values for $E_{1}$ and $M_{2}$ are presented in Tab.~\ref{tab:FF}, which we then extrapolate to the continuum limit using a simple linear $a^{2}$ fit, cf. Fig.~\ref{fig:cont_extr_F1c}. The corresponding $\chi^{2}/\mathrm{dof}\simeq 0.3$ for $E_{1}$, and $\chi^{2}/\mathrm{dof}\simeq 0.9$ for the ratio $M_{2}/E_{1}$ are satisfactory. To estimate the systematic error associated to the continuum limit extrapolation we have performed a second linear fit excluding the data at the coarsest lattice spacing. The corresponding fits are shown in Fig.~\ref{fig:cont_extr_F1c} as orange bands. Our final results, obtained by averaging the two different type of fits using the Bayesian Akaike Information Criterion\footnote{For more details on the BAIC procedure adopted, we refer to the discussion around Eqs.~(21)-(22) of Ref.~\cite{Becirevic:2025ocx}.} (BAIC) read:
\begin{figure}[!t]
\centering
\includegraphics[scale=0.37]{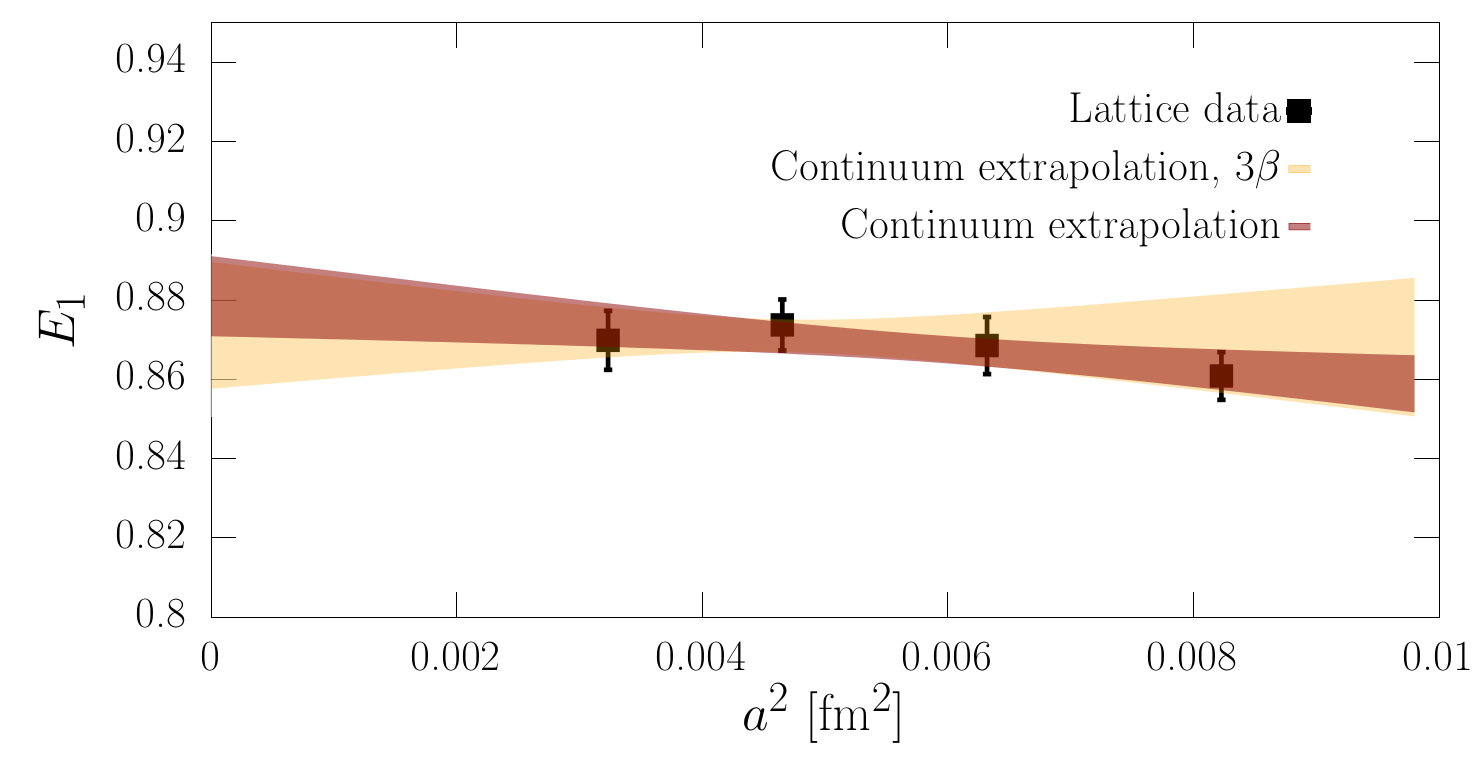}\\
\includegraphics[scale=0.37]{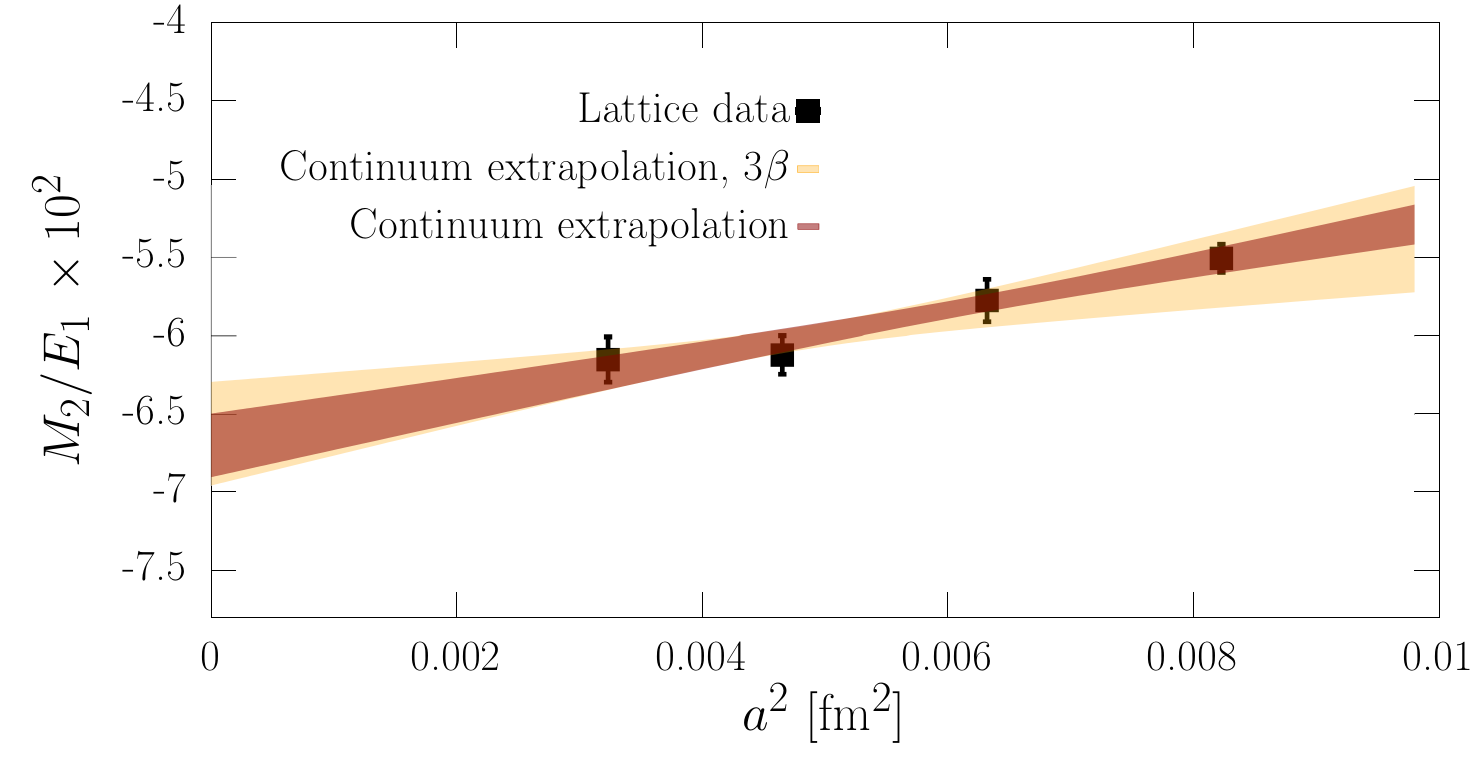}
\caption{\small\sl Continuum-limit extrapolation of the form factor $E_{1}$ (top) and of the ratio $M_{2}/E_{1}$ (bottom). The coloured bands in dark red correspond to the best-fit function obtained after performing a linear fit in $a^2$ to our lattice data. The reduced $\chi^{2}$ of the fit is about $0.3$ and $0.9$ for $E_{1}$ and $M_{2}/E_{1}$, respectively. The orange bands in the figures correspond instead to the linear fits performed excluding the data at the coarsest lattice spacing.   \label{fig:cont_extr_F1c}}
\end{figure}
\begin{align}\label{eq:FFs}
\boxed{E_{1} = 0.879(11)\,,\qquad \frac{M_{2}}{E_{1}}= -0.0668(22)\,.}
\end{align}
Note that our $E_{1}$ has an uncertainty of about $1.2\%$, while the uncertainty of the ratio $M_{2}/E_{1}$ is about 
$3\%$. We now compare our results with existing experimental data as well as with previous lattice results.

\section{Comparison with existing experimental and lattice results}
\label{sec:comparison}

After inserting our results~\eqref{eq:FFs} into Eq.~\eqref{eq:decay_rate_charm} we obtain
\begin{align}
\label{eq:final_decay_rate_charmonium}
\Gamma(\chi_{c1}\to J/\psi \ \gamma ) = 0.3265(79)~\mathrm{ MeV}\,, 
\end{align}
where we used $\alpha_\mathrm{em}^{-1} = 137.036$. The decay rate is dominated by the electric dipole form factor $E_{1}$, while the contribution of $M_{2}$ is negligible compared to $E_{1}$ at the current level 
of precision. Our result is $1.1\sigma$ larger than the experimental value $\Gamma(\chi_{c1}\to J/\psi\,\gamma)^{\rm exp} = 0.302(21) ~{\rm MeV}$, which we obtain by combining $\mathcal{B}(\chi_{c1}\to J/\psi\gamma)^{\rm exp} = (34.3\pm 1.3)\%$~\cite{PDG2024} with the PDG average $\Gamma(\chi_{c 1}) = 0.88(5)~{\rm MeV}$~\cite{PDG2024}. If, instead, we use $\Gamma(\chi_{c 1}) = 0.84(4)~{\rm MeV}$, as suggested by the global PDG fit, we would have $\Gamma(\chi_{c1}\to J/\psi\,\gamma)^{\rm exp} = 0.288(18) ~{\rm MeV}$, which is $2\sigma$
lower than our result~\eqref{eq:final_decay_rate_charmonium}. 
We emphasize that $\Gamma(\chi_{c1}\to J/\psi\,\gamma)$ has only been measured by the Fermilab E835 Collaboration~\cite{Andreotti:2005ts}. An independent measurement by BES~III would be very welcome. The difference might be also due to disconnected contributions which were omitted in our computation and will be addressed elsewhere.

As already mentioned in Introduction, the only two existing lattice QCD results have been obtained either at a single lattice spacing in the quenched approximation~\cite{Dudek:2009kk} or with two lattice spacings and by including  $N_f=2$ (unphysical) dynamical light quarks~\cite{Li:2022cfy}. With respect to $\Gamma(\chi_{c 1}\to J/\psi\, \gamma ) = 0.270(70)~\mathrm{MeV}$, reported in Ref.~\cite{Dudek:2009kk}, our result~\eqref{eq:final_decay_rate_charmonium} can be viewed as a substantial improvement. On the other hand, our result is much larger than $\Gamma(\chi_{c 1}\to J/\psi\, \gamma ) = 0.070(20)~{\rm MeV}$, obtained at finer of the two lattice spacings considered in Ref.~\cite{Li:2022cfy}. We suspect that the small decay width reported in Ref.~\cite{Li:2022cfy} is a consequence of using the TM regularization for the $\chi_{c1}$ interpolating field. As can be seen in Fig.~1 of Ref.~\cite{Li:2022cfy}, the interpolator they adopted has a clear overlap with the $J/\psi$ meson.

\begin{figure}
    \centering
    \includegraphics[width=0.9\linewidth]{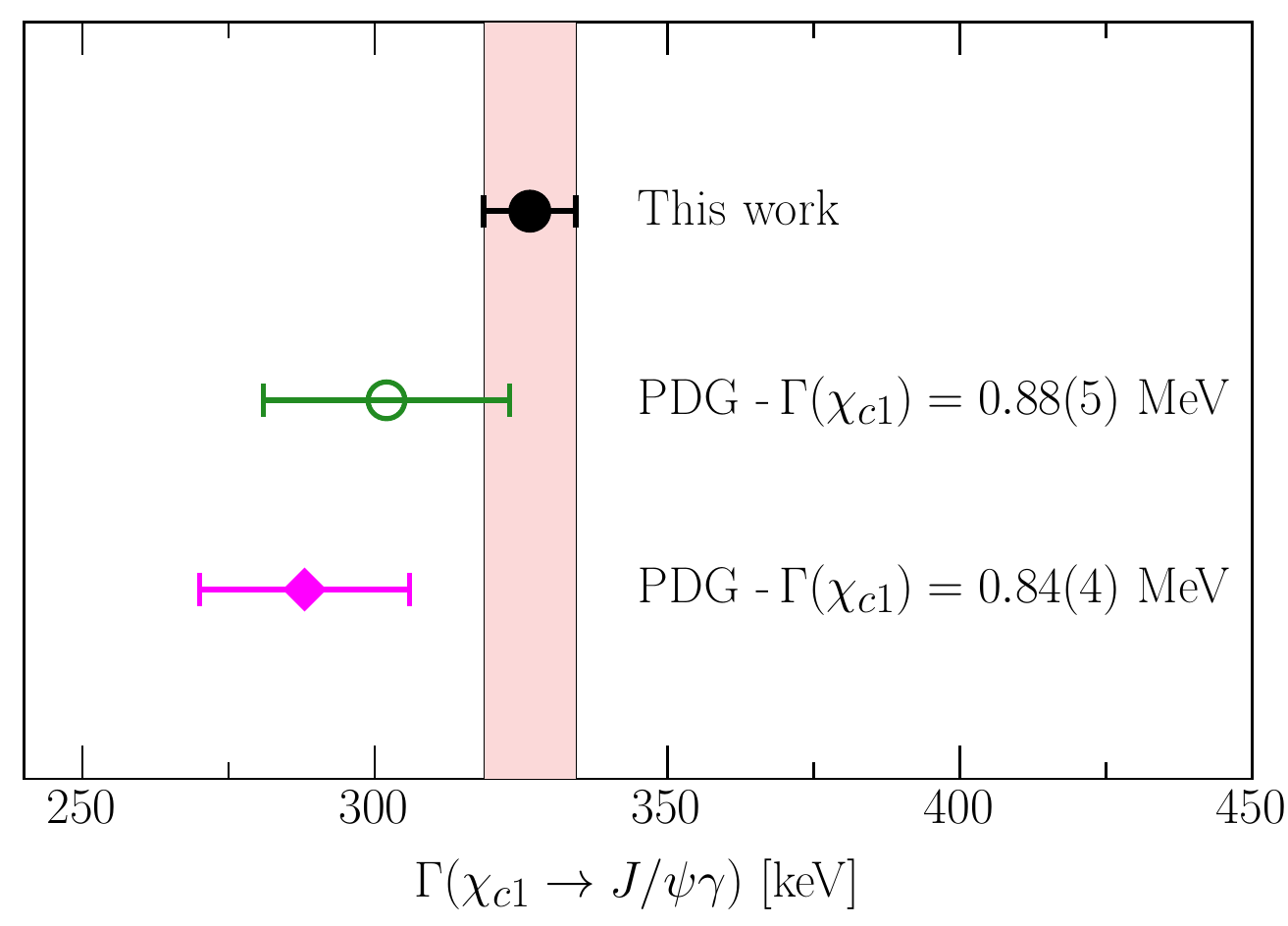}
    \includegraphics[width=0.9\linewidth]{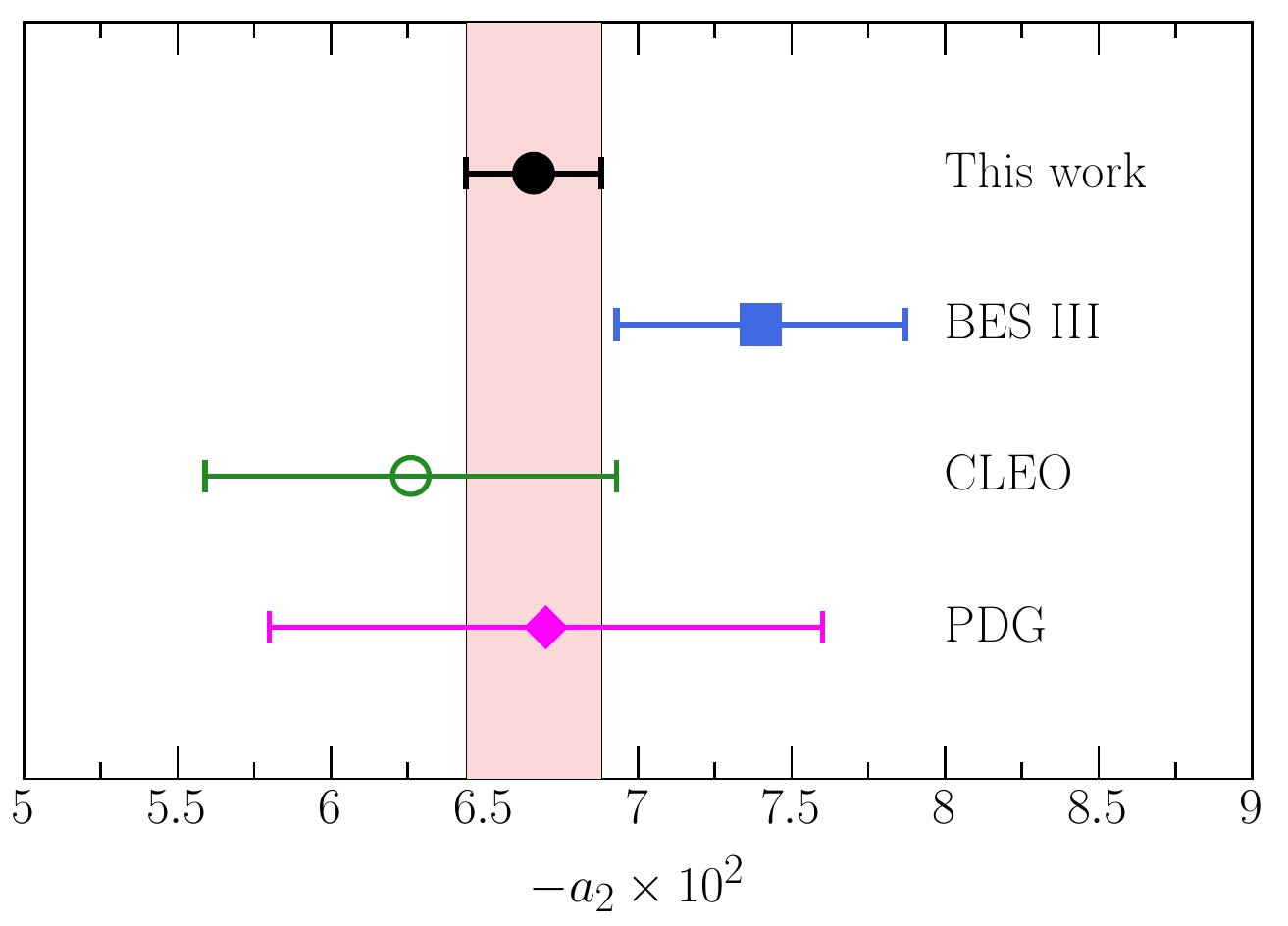}
   \caption{\small\sl Comparison between our results and experimental ones. In the upper panel our value for 
$\Gamma\bigl(\chi_{c1}\to J/\psi\,\gamma\bigr)$ (filled black circle) is shown alongside the value obtained by the PDG~\cite{PDG2024} using the  average value $\Gamma(\chi_{c1}) = 0.88(5)~{\rm MeV}$ (empty green circle) and the PDG global fit result $\Gamma(\chi_{c1}) = 0.84(4)~{\rm MeV}$ (filled magenta diamond). In the lower panel we compare our estimate of $a_{2}$ (filled black circle) with the measurement by BES~III~\cite{BESIII:2017tsq} (filled blue square) and CLEO~\cite{CLEO:2009inp} (empty green circle), as well as with the PDG average~\cite{PDG2024} (filled magenta diamond).}
\label{fig:comparison}
\end{figure}

The quantity $a_{2}$, defined in Eq.~\eqref{eq:a2}, can now easily be evaluated from our form factor ratio~\eqref{eq:FFs}. We have:
\begin{align}
    a_2=-0.0666(22)\,.
\end{align}
That same quantity is experimentally accessible through the angular distribution analysis of the decay chain $e^+e^- \to \psi(nS) \to \chi_{c 1}\gamma$ followed by $\chi_{c 1}\to J/\psi \, \gamma \to e^+e^- \gamma$, cf. Refs.~\cite{Rosner:2008ai,Sebastian:1992xq,Karl:1980wm} and Refs.~\cite{CLEO:2009inp,BESIII:2017tsq}. 
The experimental average, $a_{2}^\mathrm{exp}=-0.067(9)$~\cite{PDG2024}, is in remarkable agreement with our result~\eqref{eq:FFs}. This can be viewed as a very fine verification of the validity of the lattice QCD approach when it comes to the precision computation of hadronic quantities. 
The only other lattice QCD determination of this quantity was reported in~\cite{Dudek:2009kk}, where $a_{2}=-0.09(7)$ and $a_{2}=-0.20(6)$ were quoted, reflecting the ambiguity in extrapolating the form factors to $q^2=0$, i.e., whether they are extrapolated separately and then combined in a ratio, or whether the ratio $M_{2}(q^2)/E_{1}(q^2)$ is directly extrapolated to $q^{2}=0$. Our result for $a_{2}$ thus represents a very significant improvement in precision, by more than an order of magnitude.
In Fig.~\ref{fig:comparison} we show a comparison between our results~\eqref{eq:FFs} and the available experimental values.

\section{Conclusions}
\label{sec:conclusions}

In this paper we provide the first full QCD computation of the hadronic matrix element relevant to the $  \chi_{c 1}\to J/\psi \, \gamma$ decay by employing the gauge field configurations produced by the ETMC with $N_{f}=2+1+1$ Wilson-Clover twisted-mass quarks. A delicate point in this project was to avoid the spurious mixing with operators of the opposite parity which can plague the analysis when using the TM regularization. Using the OS regularization for the valence quarks solves this problem and the lowest lying state coupling to the $J^{PC}=1^{++}$ interpolating field operator is indeed $ \chi_{c 1}$, and not $J/\psi$ which is lighter and corresponds to $J^{PC}=1^{--}$. This is so because the OS regularization preserves the charge conjugation. We checked that feature numerically and showed that the OS valence regularization indeed provides favorable conditions for the computation of our hadronic matrix element. To ensure that the photon is on-shell ($q^2=0$) we tuned the twisted boundary conditions to one of the charm quark propagators in such a way that it actually gives a desired three-momentum to the daughter meson, $J/\psi$. In addition to the dominant electric dipole form factor $E_1$, due to the high statistics of the four gauge ensembles used in this work, we were also able to compute the ratio of the magnetic quadrupole form factor $M_2$ and $E_1$. We checked the smooth scaling of our results with the squared lattice spacing ($a^2$) and then extrapolated them to the continuum limit. 
Our final results are given in Eq.~\eqref{eq:FFs} which represent a significant improvement compared to the previous lattice studies. Our results give a decay rate that is about $(1\div 2)\sigma$ larger than the one inferred from the experimental measurements, assuming that the full width of $\chi_{c 1}$ is the one provided in Ref.~\cite{PDG2024}, which still needs to be confirmed by BES~III.

The more recent experiments focused on a detailed angular analysis of this decay, allowing one to extract $a_2$, the quantity that measures the amount of the (suppressed) magnetic quadrupole contribution to the decay amplitude, relative to the (dominant) electric dipole one. Our result is in agreement with the experimental ones, and it could be seen as strong confirmation of the validity of the lattice QCD approach when dealing with the requirement of high precision for hadronic quantities. 

Throughout this computation, we did not account for the disconnected diagrams. Being either Zweig or $\rm{SU}(3)$ suppressed, these contributions are expected to be tiny. However, this should be verified through direct computation, which is a direction we plan to explore in the near future.

\section{Acknowledgments}
\label{sec:akno}
We thank the ETMC for the most enjoyable collaboration. V.L., F.S., G.G., R.F., and N.T. are supported by the Italian Ministry
of University and Research (MUR) and the European
Union (EU) – Next Generation EU, Mission 4, Component 1, PRIN 2022, CUP F53D23001480006. 
F.S. is supported by ICSC – Centro Nazionale di Ricerca in High Performance Computing, Big Data and Quantum Computing, funded by European Union - Next Generation EU and by Italian  Ministry of University and Research (MUR) projects FIS\_00001556 and PRIN\_2022N4W8WR. We acknowledge support from the LQCD123, ENP, and SPIF Scientific Initiatives of
the Italian Nuclear Physics Institute (INFN). 
This project has received support from the IN2P3 (CNRS) Master Project HighPTflavor.

The open-source packages tmLQCD~\cite{Jansen:2009xp,Abdel-Rehim:2013wba,Deuzeman:2013xaa,Kostrzewa:2022hsv}, LEMON~\cite{Deuzeman:2011wz}, DD-$\alpha$AMG~\cite{Frommer:2013fsa,Alexandrou:2016izb,Bacchio:2017pcp,Alexandrou:2018wiv}, QPhiX~\cite{joo2016optimizing,Schrock:2015gik} and QUDA~\cite{Clark:2009wm,Babich:2011np,Clark:2016rdz} have been used in the ensemble generation.

We gratefully acknowledge the ICSC - Centro Nazionale di Ricerca in High Performance Computing for providing computing time under the allocations RAC 1916318. We gratefully acknowledge CINECA for the provision of GPU time on Leonardo supercomputing facilities under the specific initiative INFN-LQCD123, and under project IscrB VITO-QCD and project IscrB SemBD. We gratefully acknowledge EuroHPC Joint Undertaking for awarding us access to MareNostrum5 through the project EHPC-EXT-2024E01-031. The authors gratefully acknowledge the Gauss Centre for Supercomputing e.V. (www.gauss-centre.eu) for funding this project by providing computing time on the GCS Supercomputers SuperMUC-NG at Leibniz Supercomputing Centre. The authors acknowledge the Texas Advanced Computing Center (TACC) at The University of Texas at Austin for providing HPC resources (Project ID PHY21001). We gratefully acknowledge PRACE for awarding access to HAWK at HLRS within the project with Id Acid 4886. We acknowledge the Swiss National Supercomputing Centre (CSCS) and the EuroHPC Joint Undertaking for awarding this project access to the LUMI supercomputer, owned by the EuroHPC Joint Undertaking, hosted by CSC (Finland) and the LUMI consortium through the Chronos programme under project IDs CH17-CSCS-CYP. We acknowledge EuroHPC Joint Undertaking for awarding the project ID EHPC-EXT-2023E02-052 access to MareNostrum5 hosted by at the Barcelona Supercomputing Center, Spain. We also thank the GENCI.fr for granting us access to the Jean Zay computers of the computing center IDRIS in Orsay.


\bibliography{biblio}

\end{document}